\documentclass[aps, prl, reprint]{revtex4-2}
\usepackage{graphicx} 
\usepackage{bm}
\usepackage{natbib}
\usepackage{verbatim}
\usepackage{epstopdf}
\usepackage{hyperref}
\usepackage{xcolor}
\usepackage{amsmath}

\begin{document}
	
	\title{High frequency breakdown in quantum dots}
	
	\author{Nathan Johnson}
	\email[]{nathan.johnson@ucl.ac.uk
		\newline Present Address: London Centre for Nanotechnology, University College London, 17-19 Gordon Street, London, WC1H 0AH, United Kingdom}
	\affiliation{NTT Basic Research Laboratories, NTT Corporation, 3-1 Morinosato Wakamiya, Atsugi, Kanagawa, 243-0198, Japan}
	
	\author{Gento Yamahata}
	\affiliation{NTT Basic Research Laboratories, NTT Corporation, 3-1 Morinosato Wakamiya, Atsugi, Kanagawa, 243-0198, Japan}
	\author{Akira Fujiwara}
	\email[]{akira.fujiwara@ntt.com}
	\affiliation{NTT Basic Research Laboratories, NTT Corporation, 3-1 Morinosato Wakamiya, Atsugi, Kanagawa, 243-0198, Japan}

	\date{\today}
	
	\begin{abstract}
		Dynamic quantum dots are known to generate very accurate currents and can be used as a source of single electron wavepackets for use in quantum metrology, sensing or information processing. To realise their full technological potential, it is desirable to maximise the frequency of operation to increase the current. However, it has been observed that the mechanism of electron transfer across the quantum dot consistently breaks down for GHz frequencies, and this remains unexplained.
		Here, we present a novel analysis technique, combined with detailed modelling, to present a mechanism of high frequency breakdown in quantum dots as a rapidly-imparted momentum impulse. Such understanding aids future design and operation protocols, allowing their use in high frequency and real time quantum measurements and technologies.
	\end{abstract}
	\maketitle

Quantum dots (QDs) have traditionally found widespread use in studies of electron transport in condensed matter physics \cite{Kaestner3, Reilly, Kouwenhoven2} with many of these studies having focussed on utilising dc or low frequency transport.
Recently there have been proposals to utilise QDs as a high-accuracy current source, or as fast sensing devices \cite{Kaestner3, Degen}.
In these proposals, the QD is either used as a source of high fidelity wavepackets \cite{JonNatNano, Sherkunov, Bocquillon2, Bocquillon} or as a tool itself to study fast quantum coherent events such as excitation or entanglement \cite{Yamahata5, Bose, vanWiel}.
In both classes of experiment, the QD is used dynamically, with its energy with respect to the leads or its confinement being changed in time, in order to favour electron capture, ejection, select specific states or preserve coherence.
To achieve this, it is desirable to move to high frequency operation ($\sim$~GHz), either to increase the number of wavepackets created for ease of measurement \cite{Giblin, JonNatNano, Kaestner2} or to sample coherent phenomena \cite{inprep, Yamahata5, Ryu}.
Further, a high frequency QD can aid realisation of quantum computing architectures that utilise coherent QD states \cite{Koppens, Fernando, Shi}, or as high-frequency qubit structures and linkages \cite{Cai, Edlbauer, Xue}.

The state of the art QD architectures employed in such applications utilise an electrostatically defined QD embedded in a semiconductor heterostructure. 
The QD is defined by time-varying electrostatic tunnel barriers that permit electron transport into and out of the QD in specified time windows, enabling a clock-controlled charge transfer process.
In the above applications, charge may be transferred on a single, one-off process, or continuously, as a stream of individual wavepackets that form a dc current. 
In the present study, we will focus on the breakdown of the dc current in the high frequency regime, and we will show that the mechanism of this breakdown is based on single electron transfer events, and so applies to a wide range of QD use cases and material construction.

The accuracy of this charge transfer process, and hence its usability in technological applications, depends on several factors.
The detailed mechanism of the electron transfer across the QD in these electrostatic dynamic quantum dots is well understood in the adiabatic limit, when the transfer process occurs with the electron remaining in the QD ground state. 
Empirically, the adiabatic limit is found to be approx. 1~GHz in this important class of quantum dot.
The principal limitation of high frequency operation in QDs arises from the non-adiabatic (NA) excitation \cite{Giblin, Kataoka2}.
This generic term describes processes that cause unwanted excitation to a resident electron, which allows a greater chance of escape from the QD, and so the electron capture process fails and hence the accuracy of transfer decreases.
Additionally the coherence may be reduced.
This reduction in accuracy has been shown to be detrimental for the above applications \cite{Kaestner3, GiblinReview}.
Typically it is seen that the generation of dc current is seen to deviate from the adiabatic model (detailed later) and decline to zero.
Whilst observed in many electrostatically defined architectures, demonstrating the material independence and electric origins of the effect, the exact mechanism of this NA-induced breakdown remains largely unknown. 
Here we begin to clarify this by detailing a mechanism of this breakdown that can influence future gated semiconductor designs and measurement protocols and allow the realisation of high frequency quantum measurements and wavepacket sources.

In this work, we use a novel analysis technique based only upon the dc throughput current and develop a model that reveals the NA excitation to be a sharp momentum transfer imparted to an electron resident in the QD.
In our study, we use a gate-defined silicon QD and operate it dynamically \cite{Giblin, Fujiwara, Fujiwara2, Rossi}, such that a well defined number of electrons can be loaded into the QD and expelled with a recurring frequency $f$.
Our system is analogous to those constructed in other semiconductor media \cite{GiblinReview, Kaestner3}.
We utilise the well-verified adiabatic model as a basis of our study of the NA excitation \cite{Slava}.
In the NA case, some electrons are excited on loading the dynamic QD, which will induce further electron escape back to source from this excited state.
In our work, we utilize a higher-energy (HE) state to sample these excited electrons, which can be either a trap sate \cite{Yamahata2, Yamahata5, Fujiwara3} or an intrinsic excited state of the QD.
We will refer to this state as the higher-energy (HE) state throughout this work. 
Additionally, our method provides a complimentary technique to study the phonon coupling to a resident electron, as we also account for thermal activation of the HE state.
We derive a model that can recreate the thermal and NA components of the excitation with considerable accuracy and find that the NA-excitation can only be understood as a sudden impulse on the electron, whereas the thermal excitation can be separated as a lattice phonon interaction as expected.
Our novel analysis technique of deriving excitation rates from dc current could prove useful in understanding and constructing future quantum measurements and enhance sensing protocols, in addition to the knowledge gained of the origin of the excitation itself.
  
 \section{Measurement of Non-Adiabatic excitation}
 
 \begin{figure}
 	\includegraphics[width=\linewidth]{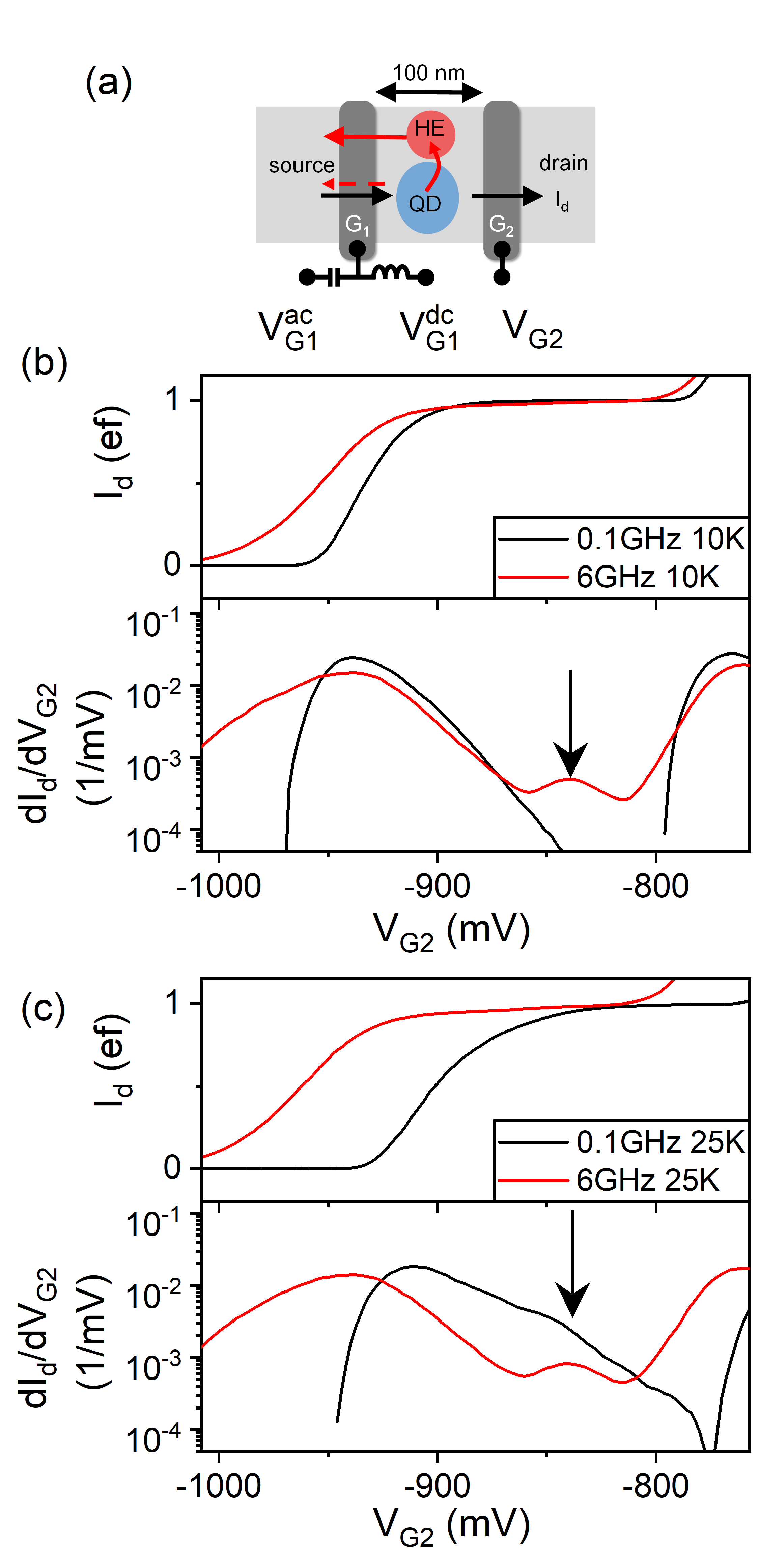}
 	\caption{
 		(a) Schematic of the device used in this work. The main QD is shaded in blue, and the HE state in red. Arrows denote the possible electron paths: black, capture and forward escape, defining $I_{d}$; red solid, escape to source via the HE state; red dash, direct escape to source.
 		(b), (c) Upper panel: QD current $I_{d}$ shows a clear plateau reflecting the capture process. Lower panel: the derivative $dI_{d}/dV_{G2}$ shows the captured electron distribution. Black trace, 0.1~GHz, adiabatic capture; red, 6~GHz, NA excitation present. Arrow highlights the distribution formed from the HE state population. At 10~K, (b) shows no HE state component at 0.1~GHz, but in (c) at 25~K, we see thermal activation to the HE state can occur. 
 	\label{Fig1}
 	}
 \end{figure}	

A sketch of the device used is shown in Fig.~\ref{Fig1}(a). 
The QD, shaded in blue, is formed electrostatically between two polycrystalline-Si gates $\rm{G_{1}}$ and $\rm{G_{2}}$, separated by 100~nm, that lie above a Si-nanowire \cite{Johnson3}.
An additional upper gate covers the area in the sketch to aid conduction.
The QD is formed when we apply the potentials $V_{G1}$ and $V_{G2}$ to the gates $\rm{G_{1}}$ and $\rm{G_{2}}$ respectively.
We operate the QD dynamically, in a similar way to single-electron current sources \cite{Fujiwara2, Kaestner3}.
$V_{G1}$ is driven periodically by a sinusoidal waveform $V_{G1}^{ac}(t)$ with period $1/f$, whilst $V_{G2}$ is held constant.
During a single period of $V_{G1}^{ac}(t)$, the potential underneath $\rm{G_{1}}$ drops below the Fermi energy, allowing electrons from the source to populate the QD region. 
As the potential rises, the QD is formed and electrons are isolated within it. 
As the potential continues to rise, the QD potential rises above that of $V_{G2}$, and the resident electrons are ejected to the drain.
This forms a dc current $I_{d} = nef$, with $n$ the captured population of the QD and $e$ the electronic charge.
This process is denoted with black arrows in Fig.~\ref{Fig1}(a).
We develop our analysis using only a straightforward measurement of $I_{d}$.
Because the ejection process from the QD can be made to occur with unit probability, $I_{d}$ is a measure of the capture dynamics, and hence this capture stage determines the effectiveness of the QD to act as a current source or of coherent wavepackets.
This electrostatic scheme is material independent \cite{GiblinReview, Kaestner3} and allows us to measure the generic incoming process to a QD.
A definition of accuracy of this process follows naturally by comparing $I_{d}$ to the expected current for capturing $n$ electrons per cycle $1/f$ with certainty.
The sample is mounted in a temperature-controllable dilution refrigerator in zero magnetic field.

In our device, we also observe the presence of a HE state, which is known to occur in silicon devices \cite{Yamahata2, Yamahata5}.
We speculate this state could be an instance of two possibilities: it could be an intrinsic excited state of the QD, or it could be a (trap) state of a parasitic QD in parallel to the main QD, as depicted in Fig.~\ref{Fig1}(a).
In the case of the HE state being attributable to a trap-like QD, we find in our case its ground state energy is at a higher potential than that of the main QD, as we do not see parallel current flow across the QD system.
The HE state serves to return electrons to the source lead, depicted with the arrows in Fig.~\ref{Fig1}(a), as we will show later.

\begin{figure}
	\includegraphics[width=\linewidth]{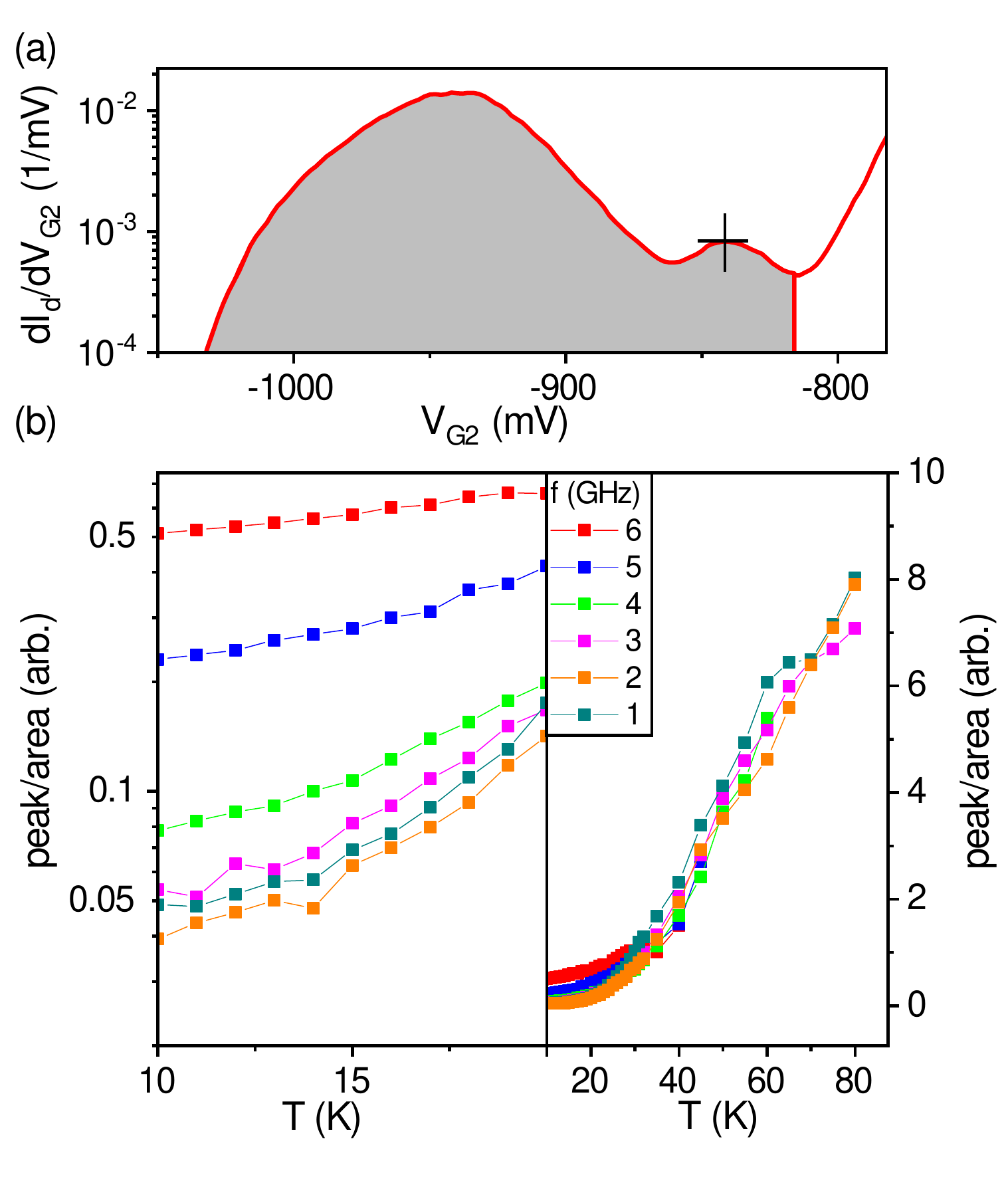}
	\caption{
		(a) Example of the quantitative sampling peak-to-area method. The peak is marked with a cross and the shaded area denotes the captured population.
		(b) The peak-to-area measure of the HE state population. Left panel: For $f = 1 - 6$~GHz we see a progression in the ratio attributable to NA excitation. Right panel: Temperature evolution at all frequencies. 
	\label{Fig2}
	}
\end{figure}

In Figs.~\ref{Fig1}(b) and \ref{Fig1}(c) we show the current $I_{d}$ (upper panel) and the derivative $dI_{d}/dV_{G2}$ (lower panel).
As $V_{G2}$ is varied from higher potential (more negative voltage) to lower, plateaus of current are seen in $I_{d}$ for successive $n$.
In this study, we restrict ourselves to the first plateau, where $n = 1$, which corresponds to ejection of a single electron from the QD, and is of greatest interest in most applications. 
Our analysis extends to multiple $n$ due to the nature of charge transfer and excitation being based on single electron processes \cite{Slava}.
We use a frequency $f$ of 0.1~GHz as a basis for adiabatic transfer through the QD, plotted in black in Figs.~\ref{Fig1}(b) and \ref{Fig1}(c).
At this frequency, electron transport across the QD occurs adiabatically \cite{Slava, Kaestner3, Brandes} and a dynamic model of the capture process has been verified \cite{Slava, Kaestner, Yamahata, Yamahata3}.
This model, commonly referred to as the decay cascade model, develops the time dependence of the capture process and has been found to predict $I_{d}$ to very high accuracy \cite{GiblinReview} within the adiabatic limit.
In our device, we also confirm the accuracy of this model for $I_{d}$ at $f = 0.1$~GHz. 
We use this confirmation as a basis to verify that $I_{d}$ measured at $f = 0.1$~GHz is entirely developed from adiabatic (ground state) capture to the QD.
We will use this adiabatic data set as a basis to compare the ground state and excited cases.

The derivative $dI_{d}/dV_{G2}$ shows a well-defined characteristic distribution reflecting the capture statistics, and is plotted in Fig.~\ref{Fig1}(b) (lower panel, black trace).
By contrast, we plot in red the equivalent trace at $f=6$~GHz.
Here, we see this distribution is broadened, which we attribute to the high-frequency NA breakdown in the capture process, and hence the accuracy of $I_{d}$ is reduced.
Additionally, highlighted with an arrow in Fig.~\ref{Fig1}(b), we see a separate distribution which we attribute to electrons being promoted to the HE state by the NA excitation, and returning to source (see Fig.~\ref{Fig1}(a)) \cite{Kataoka2}.
The HE state is thus sampling a proportion of QD electrons that probabilistically undergo the NA excitation process.
The HE state distribution is equivalent to a population of excited electrons.
By studying this distribution under frequency and temperature (which principally drive the NA and phononic components respectively) we can derive a model of the NA excitation process.

In Fig.~\ref{Fig1}(b), the traces are recorded at a temperature of $T=10$~K, which is sufficiently low enough to maintain ground state capture, and thermal fluctuations are negligible \cite{Johnson3, Yamahata3, Fujiwarabook}.

In Fig.~\ref{Fig1}(c), we plot the equivalent measurements of $I_{d}$ but at $T = 25$~K.
We see at this elevated temperature that the HE state can be populated by the thermal activation at low $f$ (lower panel, black, highlighted).
At high $f$, there is an increase in amplitude of the HE state population.
As with NA excitation, this thermal escape to the source can be direct from the main QD also.
Both the NA excitation and thermal energy produce similar effects on the distributions but from different origins.
We will account for the electron-phonon coupling during this capture phase in addition to the NA excitation.
Hence, we see excitation giving escape to the source directly, or HE state-mediated, from two mechanisms: thermal activation or the NA excitation.
The HE state can sample this excitation by capturing these higher energy electrons.

To quantify these effects, we measure $I_{d}$ over 0.1~-~6~GHz, to see the progression of the NA breakdown, and for a temperature range $T = 10-80$~K, to measure the thermal effect.
We find that, even in the high-$f$ case where the HE state has highest occupation, the current lost via excitation to the HE state and relaxation to the source is $\sim$~pA, making it difficult to evaluate quantitatively.

We analyse the HE state population as shown in Fig.~\ref{Fig2}(a).
Here, we show a sample derivative trace $dI_{d}/dV_{G2}$ (red), similar to those of Fig.~\ref{Fig1}(b,c). The shaded area corresponds to the captured electron distribution. 
We note the similarity in shapes of the main QD and HE state distributions.
The shape is largely determined by the cross-coupling between the QD and $V_{G1}^{ac}(t)$.
Hence, the similarity between the distribution shapes allows us to infer that the cross-coupling between the HE state and $V_{G1}^{ac}(t)$ are quite similar.
We see that $I_{d} < 1$ over this region always, which confirms that the HE state is not contributing electrons to the main QD.
This allows us to assume the HE state distribution is monotonic with population.
Therefore we can take the peak height as a less noisy measure than the HE state peak area, as marked with a cross on Fig.~\ref{Fig2}(a).
In Fig.~\ref{Fig2}(b) we plot this peak-to-area ratio.
In the left panel, we can see a clear frequency dependence, with higher $f$ showing a larger HE state population, attributable to the NA excitation.
In the right panel, we can see a clear $T$ dependence, as the HE state is thermally populated, universally at all frequencies.
We note that the main distribution also varies with $T$ and $f$, likely due to escape directly to source from the main QD, and this methodology allows a natural normalisation to the total number of captured charges.

\begin{figure}
	\includegraphics[width=\linewidth]{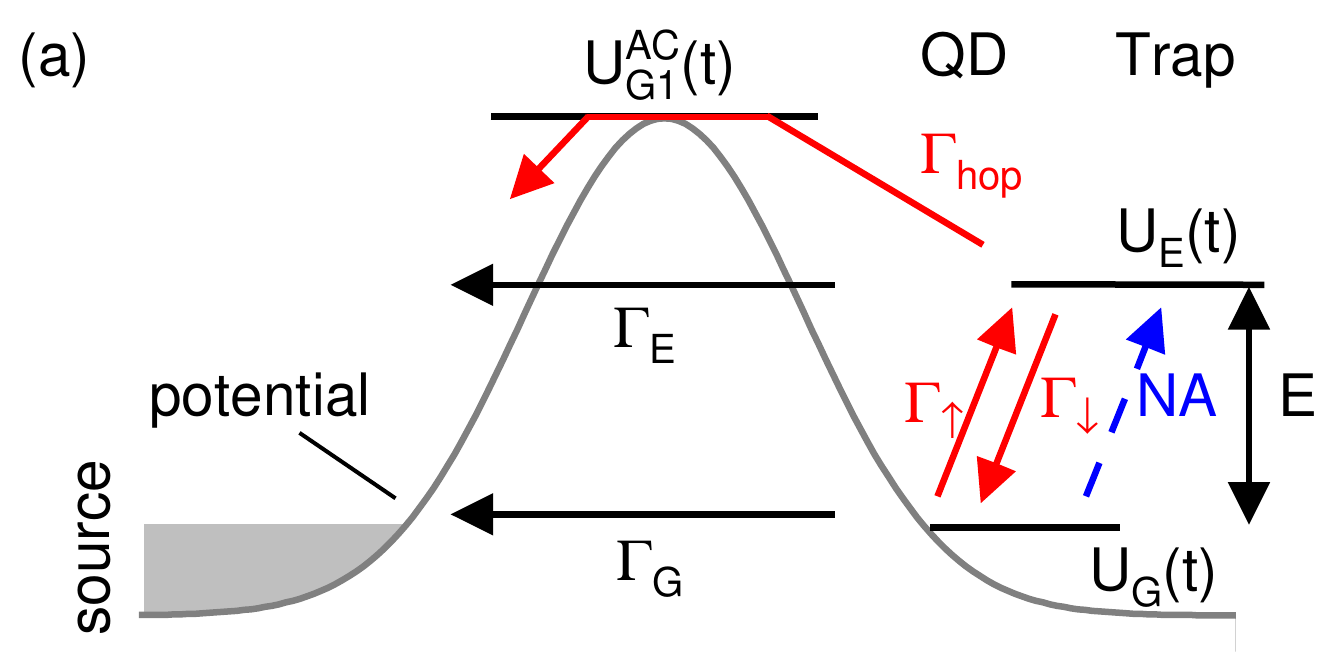}
	\includegraphics[width=\linewidth]{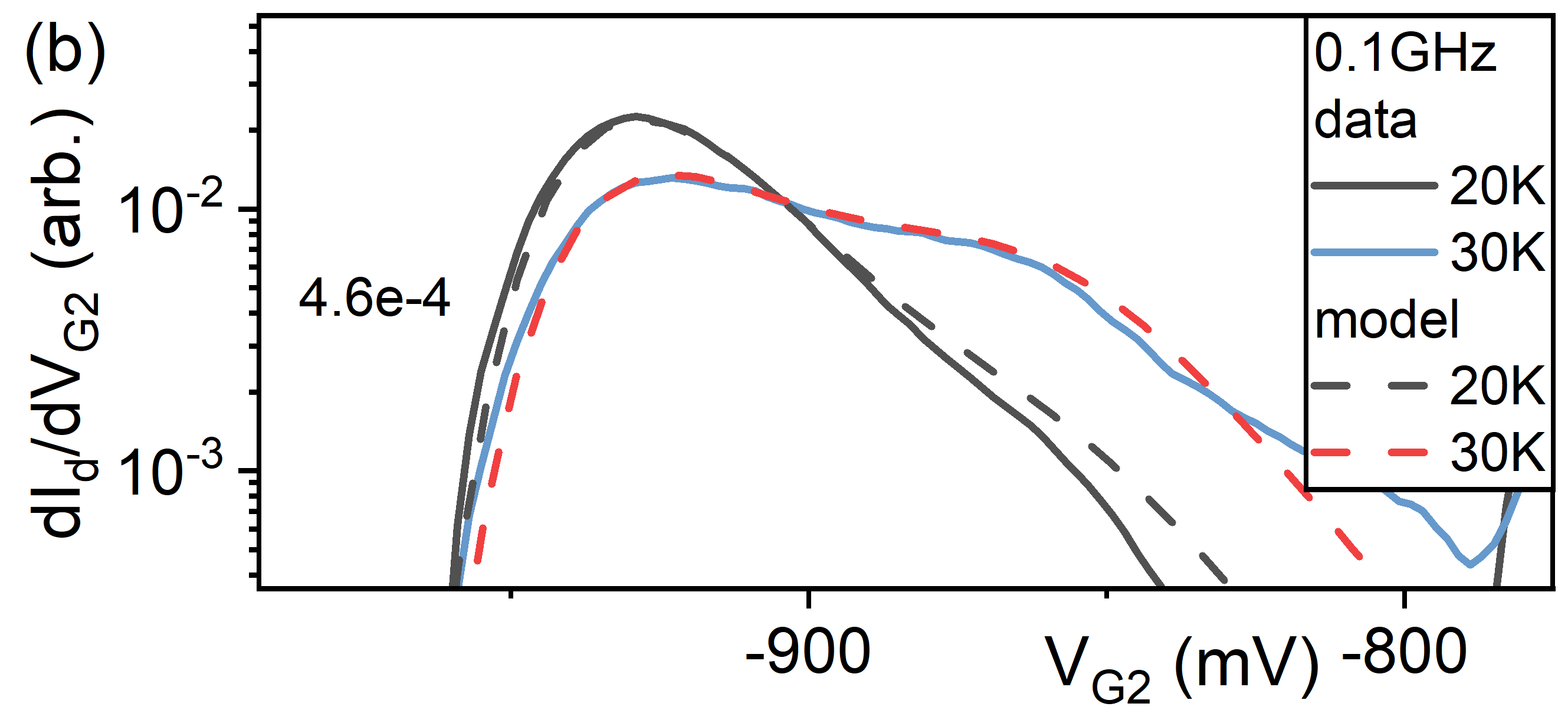}
	\includegraphics[width=\linewidth]{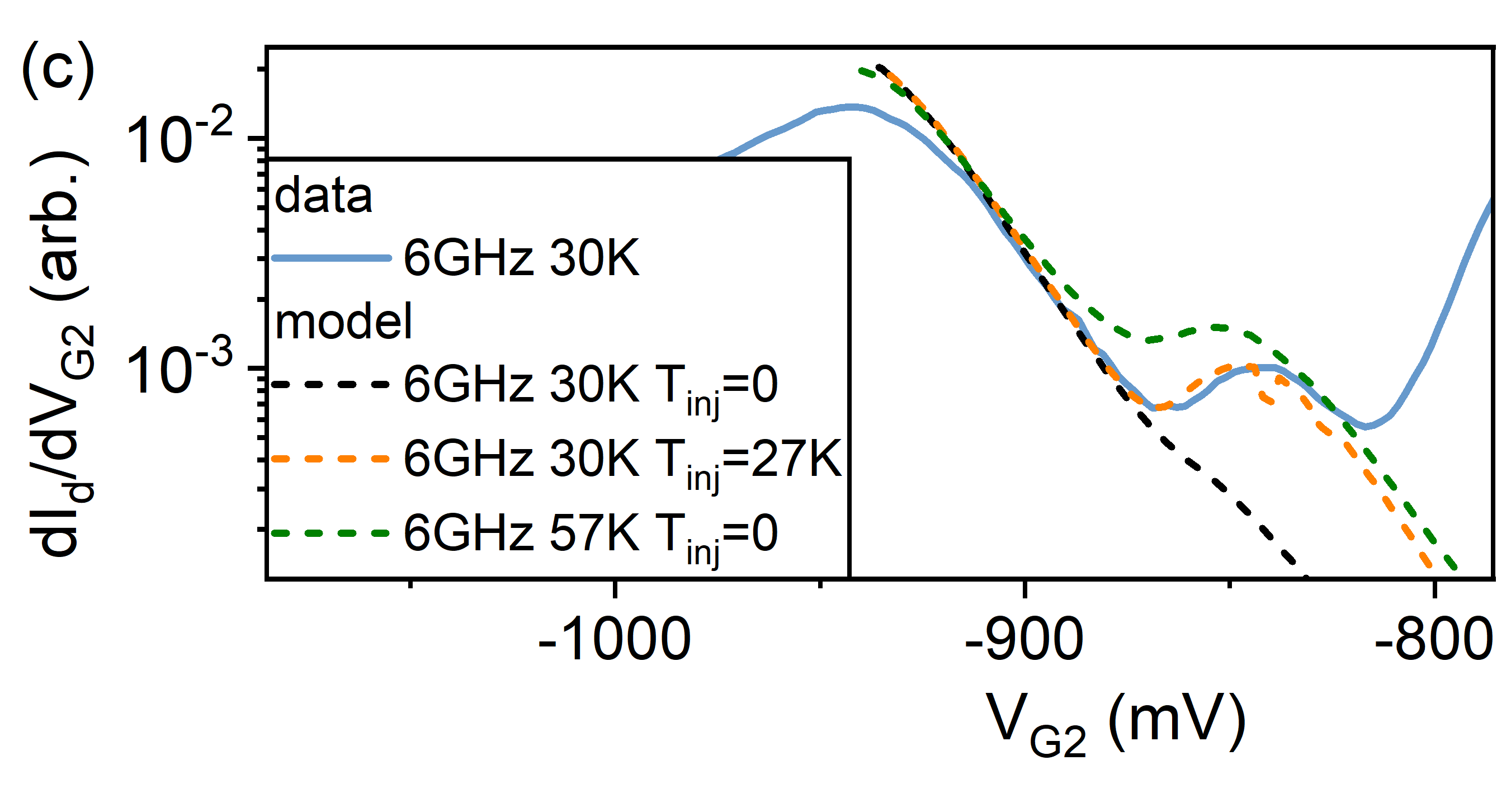}
	\includegraphics[width=\linewidth]{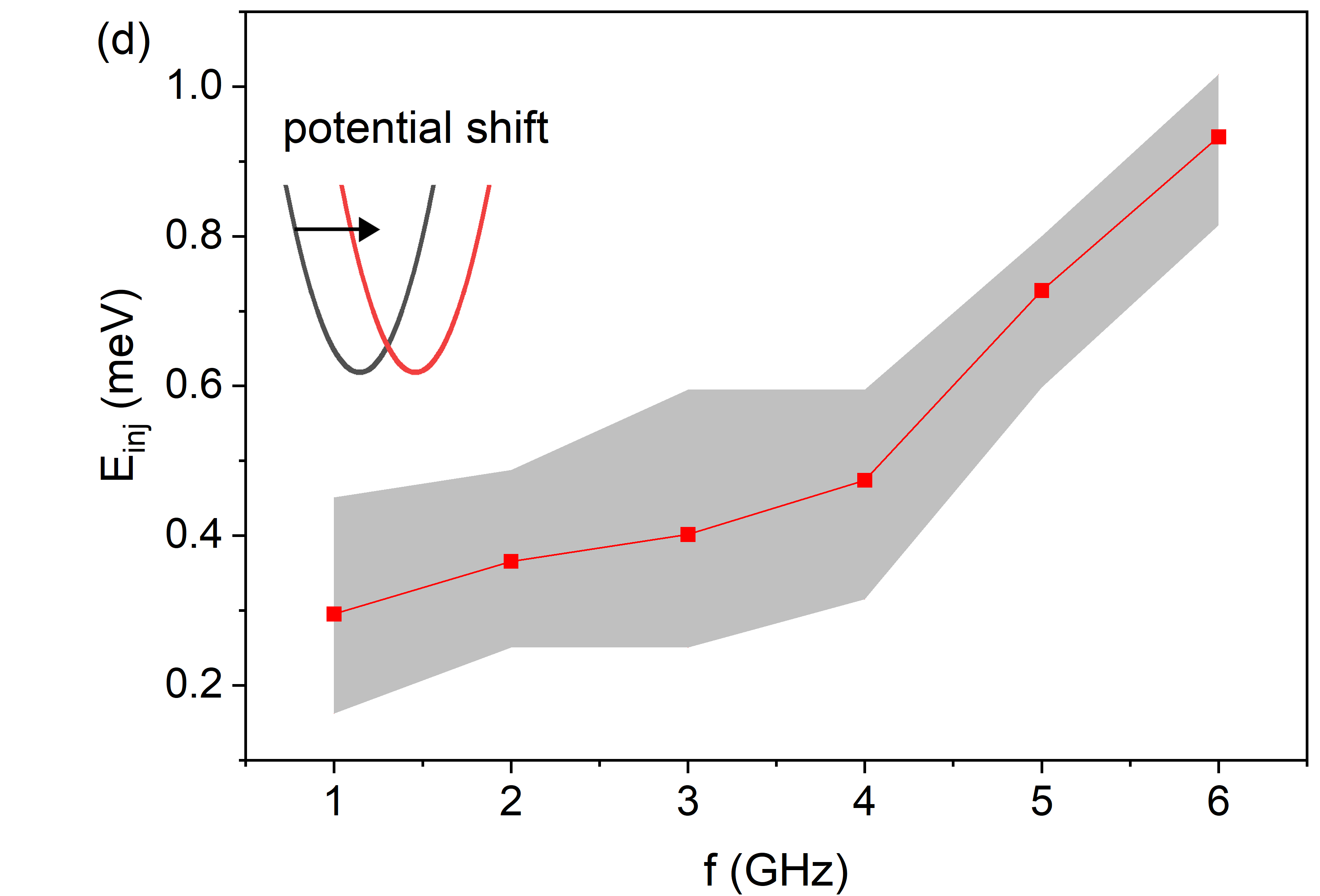}
	\caption{
		(a) Schematic of the model. Arrows denote the allowed transitions, labelled by the attempt frequency of the transition. Terms in Eqns.~\ref{eqn1} and \ref{eqn2} in the main text describe the rates of each transition. 
		(b) Determination of the pair $(\Gamma_{ph}, E)$ by mapping to 0.1~GHz (adiabatic) data finds ($1 \times 10^{11}$~$\rm{s}^{-1}$, 9~meV).
		(c) Verification of the shape of the HE state distribution. A non-zero $E_{inj}$ is required to get the peak shape of the NA distribution, whereas using an increased (lattice temperature) $T$ gives the more rounded shape, as seen also in (b). 
		(d) $E_{inj}$ with frequency $f$. The shaded region shows the variability in the range of $T = 20~-~30$~K. Sketch highlights the dominant component of $E_{inj}$ as a potential shift.
	\label{Fig3}
	}
\end{figure}

\section{Model}
To separate the thermal and NA components we derive a model as sketched in Fig.~\ref{Fig3}(a).
We construct a two level system, consisting of the QD ground state $U_{G}(t)$ and the HE state $U_{E}(t)$, which due to the similar coupling between the QD and HE state with $V_{G1}^{ac}(t)$, can be separated by a fixed energy $E$.
These two levels are used to represent the total spectrum of states available to an electron in the QD and HE state respectively.
The barrier top, denoted by $U_{G1}^{ac}(t)$, rises with $V_{G1}^{ac}(t)$. 
$U_{G}(t)$ and $U_{E}(t)$ rise together also, but at some slower rate, determined by the cross-coupling between the gate and the QD/HE state system. \cite{Yamahata5, Johnson3}.
Arrows denote the allowed transitions.
The black arrows show escape to the source, red arrows denote thermal excitation and relaxation, and the blue arrow the NA excitation. 

The time evolution of the total population of the QD/HE state system is \cite{inprep}

\begin{multline}
\frac{{\rm{d}}P_{G}(E,t)}{{\rm{d}}t}=-\Gamma_{\uparrow}P_{G}(E,t)+\Gamma_{\downarrow}P_{E}(E,t) \\ -\Gamma_{G}P_{G}(E,t)\left(1-F \left( U_{G}(t)\right)\right)T_{G}(E);
\label{eqn1}
\end{multline}

\begin{multline}
\frac{{\rm{d}}P_{E}(E,t)}{{\rm{d}}t}=\Gamma_{\uparrow}P_{G}(E,t)-\Gamma_{\downarrow}P_{E}(E,t) 
\\ -\Gamma_{E}P_{E}(E,t)\left(1-F \left( U_{E}(t)\right)\right)T_{E}^{tn}(E)
\\ -\Gamma_{hop}P_{E}(E,t)\left(1-F \left( U_{G1}^{ac}(t)\right)\right)T_{E}^{th}(E)
\label{eqn2}
\end{multline}
where $P_{G,E}(E,t)$ denotes the population $P$ of the state $U_{G,E}(t)$.
Each of the arrows depicting allowed transitions in Fig.~\ref{Fig3}(a) is labelled by the corresponding attempt frequency.

Thermal interactions are expressed as
$\Gamma_{\uparrow} = \Gamma_{ph}\exp(-E/kT)$ and $\Gamma_{\downarrow} = \Gamma_{ph}$ and follow the detailed balance, and $\Gamma_{hop}$ is a thermal escape rate.
$\Gamma_{G,E}$ is an attempt frequency from each state, and we take $\Gamma_{G} = \Gamma_{E}$; $F(\epsilon)$ is the Fermi function and $T_{G,E}(E)$ is a transmission coefficient (see the appendix for a full description of the model terms) \cite{Johnson3}.
$U_{G}(0) \equiv -E_{c}$ with $E_{c}$ the charging energy and $U_{G1}^{ac}(0)$ is variable proportionally to $V_{G2}$ \cite{Johnson3}.
$k$ is Boltzmann's constant and $T_{0}$ is a constant.
This evolution allows only for direct (tunnelling) escape from the QD/HE state and thermal activation of the HE state (see also Fig.~\ref{Fig1}(a)).
Note that we use $\Gamma_{hop} = \Gamma_{ph}$ for the thermal activation from the HE state for simplicity because the corresponding barrier height is low enough.


Fig.~\ref{Fig3}(b) plots two traces at 0.1~GHz at a low (20~K) and high (30~K) temperature. 
The data shows the HE state is populated more significantly at 30~K.
The dashed lines show the result of the above model, with no NA excitation included, where we have have used $\Gamma_{G}  = 10$~THz, $\Gamma_{ph} = 0.1$~THz and $E = 9$~meV chosen as a best fit to the data.
We find $\Gamma_{ph}$ is in agreement with literature values \cite{Dur}.

We now incorporate the NA excitation into our model.
In Fig.~\ref{Fig3}(b), the initial zero-time loading condition was a Fermi distribution shared between $U_{G}(0)$ and $U_{E}(0)$.
We introduce the NA excitation as a rapid impulse - an electron in the QD receives a sudden momentum gain which can cause the excitation.
We define $T_{inj}$ such that the zero time Fermi distribution is calculated according to temperature $T_{start} = T + T_{inj}$.
Note we are not asserting that $T_{start}$ is the real electron temperature after receiving the impulse; we are projecting a Fermi distribution onto $U_{G}(t)$, $U_{E}(t)$ and so it is an equivalent adiabatic temperature. 
We assert no requirement for the true distribution to follow the Fermi (equilibrium) distribution at these short timescales.
In our model, we only modify the zero-time populations (see the appendix for detail) $P_{G,E}(0)$ by using $T_{start}$. 
Then the dynamics proceed using only the lattice temperature $T$ for all times $t > 0$.
Fig.~\ref{Fig3}(c) compares this model with data.
We plot data for $f = 6$~GHz at 30~K.
In black dash, we show the model result using $T_{inj} = 0$, that is $T_{start} = T = 30$~K, which shows a small HE state population but far less than the data.
In orange dash, we see that we can well match the data if we choose $T_{inj} = 27$~K, and run the model with $T_{start} = T + T_{inj} = 30 + 27$~K.
By contrast, in green we also plot the effect of $T_{inj} = 0$ but $T = 57$~K, so $T_{start} = T + T_{inj} = 57 + 0$~K (effect of continuous energy injection), which clearly shows a different shape, similar to the low frequency plots of Fig.~\ref{Fig3}(b).
We see that the effect of a sudden impulse via $T_{inj}$ is to give a more rounded, well defined distribution and continual thermal activation makes more of a flatter shape, and within the context of the two-level system model, supports the conclusion of a sudden momentum impulse.
We suggest this gives strong evidence that the NA mechanism is of the form of a fast impulse rather than a continual energy injection, in agreement with a related study \cite{Yamahata5}.

$T_{inj}$ is a fit parameter required in the above model. 
To be more accurate, we define an injected energy $E_{inj}$, which better quantifies the NA impulse as an energy injection.
The condition $E_{inj} = T_{inj}$ is only true if at all timescales the QD population is in thermal equilibrium and the QD energy spectrum is a quasi-continuum, and thus well described by Fermi statistics.
There is no requirement for this to be the case, and our data does not provide any insight on this issue. 
To be fully general and quantitative, we can assert $P_{E}^{inj} = P_{E}(T+T_{inj}, t = 0) - P_{E}(T, t= 0)$, where we are separating the proportion of the initial populations that we require to match the data from the thermal equilibrium case.
Then $E_{inj} = E P_{E}^{inj}$, with $E = 9$~meV as found earlier.

In Fig.~\ref{Fig3}(d) we plot $E_{inj}$, derived by inspection of each curve for $f = 1 - 6$~GHz for $T = 20 - 30$~K in 1~K increments. 
The red trace plots the mode of the values, and the shaded area shows the range of $E_{inj}$, and clearly supports the data of Fig.~\ref{Fig2}(b).

\section{Discussion}
We interpret $E_{inj}$ as an excess kinetic energy imparted to the electron as the QD forms and is isolated from the source when $U_{G1}^{ac}(t)$ crosses the Fermi energy.
We can attribute the main mechanism for this to the dynamic shift in the position of the QD, which results in velocity gain and spatial oscillation of the SE wave packet inside the QD \cite{Yamahata5}.
There may be some smaller contribution from a changing QD confinement during the capture process.


For a potential shift, we suggest $E_{inj} \equiv \frac{1}{2}mv^{2}$, where $m$ is the reduced mass and $v$ is an excess speed of the wavepacket.
For the range $f = 1 - 6$~GHz, we find  $v = 20 - 40$~nm/ps, which is aligned with a previous estimate \cite{Yamahata5}.
We can roughly deduce the QD centre shifts by $x \sim 26 -44$~nm during the NA impulse, which is plausible with the lithographic separation of the gates of 100~nm \cite{Yamahata5}.
We expect $v \propto f$ and so $E_{inj} \propto f^{2}$, which we see some evidence of in Fig.~\ref{Fig3}(d), as we highlight with the sketch.
We note that in the limit that the QD shape remains harmonic throughout the NA process, we would expect that the excitation would result in only a slightly perturbed Fermi distribution, and thus reinforcing the $E_{inj}$ found here, although we note the caveat on the reliability of $T_{inj}$ to equate to a real temperature.

Although in general there can be another type of NA excitation from confinement-shape change, our recent results using a QD device simulator \cite{Fujiwara4} suggest that the effect would be minor because the confinement change is small and rather symmetric. 
Furthermore, the QD confinement may be likely pinned by impurities and barrier potential fluctuations in the channel, which more weakly couple capacitatively to $V_{G1}^{ac}(t)$.
In agreement with a previous study \cite{Yamahata5}, we suggest the origin of NA excitation in this case to be largely dominated by a sudden potential shift.


In conclusion, we have shown a novel analysis technique that uses only the dc throughput current of a QD that allows us to estimate the phonon coupling and NA components, and provide some insight into the time dependence of the QD potential profile.
Whilst the high frequency breakdown was often observed, direct quantitative study of its mechanism was previously elusive. 
Our method could also be adapted to provide quantum sensing of the electron wavefunction, or track dynamic changes in electric field.
We have provided a quantitative study of the mechanism of the NA breakdown in QDs, and have shown evidence to support the QD position shift to be a dominant component in the high frequency breakdown.
To overcome the breakdown, we recommend the use of a plunger gate over the QD, which can allow finer control of the QD profile and decouple it from the ac potential oscillation.
A split gate could be used as the loading gate $\rm{G_{1}}$, which could help restrict the momentum transfer to the electron.

This work was partly supported by JSPS KAKENHI Grant Number JP18H05258.

\section{Appendix: Derivation of model terms}
Here we describe the selection of parameters included in the model used to derive the NA and thermal components in the main text.
The derivation of this model and parameters largely follow discussions published in refs.~\onlinecite{Johnson3, Yamahata5, Fujiwarabook} and we only outline parameters here.
The model is described by equations~\ref{eqn1} and \ref{eqn2} in the main text.
We write the transmission coefficients used in eqns.~1 and 2 as \cite{Johnson3}
\begin{equation}
	T_{G}=\frac{\exp(-(U_{G1}^{ac}(t)-U_{G}(t))/kT_{0})}{1+\exp(-(U_{G1}^{ac}(t)-U_{G}(t))/kT_{0})}
	\label{eqn3}
\end{equation}
\begin{equation}
	T_{E}^{tn} = \frac{\exp(-(U_{G1}^{ac}(t)-U_{E}(t))/kT_{0})}{1+\exp(-(U_{G1}^{ac}(t)-U_{E}(t))/kT_{0})} 
\end{equation} 
\begin{equation}
	T_{E}^{th} = \exp\left(\frac{-(U_{G1}^{ac}(t)-U_{E}(t))}{kT}\right)
\end{equation}
Here, we have divided the transmission coefficient into two terms, corresponding to tunnelling and thermal activation respectively, and $T_{0}$ is a constant \cite{Johnson3}.
In the QD coefficient, $T_{G}$ (eqn.~\ref{eqn3}), we neglect the thermal activation term for simplicity, because there is little effect, owing to the QD having a lower potential than the 
HE state and the fast rising barrier $U_{G1}(t)$ \cite{Johnson3,inprep}. 
We use the same transmission coefficients to describe the incoming terms (the reverse processes for each of the processes that allow escape to source, see Fig.~\ref{Fig3}(a)), which are included in the analytical model. 
These terms are not written explicitly in eqns.~\ref{eqn1} and \ref{eqn2}, owing to their similarity, and they do not alter the populations significantly, owing to the NA process being a time-zero effect.
We describe the motion of the dynamic barrier potential as

\begin{equation}
	U_{G1}^{ac}(t) = 2V_{G1pk}^{ac}\alpha_{G1}ft+\alpha_{G2}V_{G2}
\end{equation}
where $V_{G1pk}^{ac}$ is the peak-to-peak amplitude of $V_{G1}^{ac} = 1.6$~V.	$\alpha_{G1}$ describes the coupling between $V_{G1}^{ac}(t)$ and $U_{G1}^{ac}(t)$, and $\alpha_{G2}$ describes the coupling between $V_{G2}$ and $U_{G}(t)$.
$\alpha_{G1} = 0.55$ is determined by examining the conductance in the 2D parameter range $V_{G1}^{ac}(t), V_{G2}$ \cite{Yamahata5}.
$\alpha_{G2} = 0.10$ is chosen by a fit of the low frequency, low temperature model result to the data.
Neither value changes with frequency \cite{Yamahata5, Johnson3}.
The QD potential rises as 

\begin{equation}
	U_{G}(t)=(\alpha_{G1-G}/\alpha_{G1})(U_{G1}^{ac}(t)-\alpha_{G2}V_{G2})-E_{c}
\end{equation}
where $\alpha_{G1-G} = g\alpha_{G1}/(1+g) = 0.5$ is the cross-coupling between $V_{G1}^{ac}(t)$ and $U_{G}(t)$.
$g$ is a measure of the coupling between the QD and $V_{G1}^{ac}(t)$ and we use $g = 10$ because of the high degree of cross-coupling \cite{Yamahata6}.
$E_{c} = 17$~meV is the charging energy and is chosen in conjunction with $T_{0} = 17$~K by the method detailed in Ref.~\cite{Johnson3}.
Hence, $U_{E}(t)$ evolves as $U_{G}(t) +E$.
In the absence of NA excitation (low frequencies), at $t=0$, $P_{G,E}(0)$ is determined as a Fermi distribution between the two states $U_{G,E}(0)$

\begin{equation}
	P_{G}(0) = F(-E_{c})\times F(-E), P_{E}(0)=F(-E_{c}) \times F(E)
\end{equation}
and note that the probability of the QD/HE state system being empty is $P_{zero} = 1-F(-E_{c})$.
This is constructed such that only the QD can load. 
Then $P_{E}(0)$ represents an initial thermal excitation taking place before the QD has completely isolated from the source.

\bibliography{reference_library}

\end{document}